\documentclass[aps,twocolumn]{revtex4}

\usepackage{graphicx}
\usepackage{latexsym}
\usepackage{graphicx}
\usepackage{epsfig}
\usepackage[english]{babel}
\usepackage[latin1] {inputenc}
\usepackage{amsmath,amsfonts,amssymb}
\usepackage{bm}

\newcommand \be {\begin{equation}}
\newcommand \bea {\begin{eqnarray}}
\newcommand \ee {\end{equation}}
\newcommand \eea {\end{eqnarray}}
\newcommand \bed {\begin{displaymath}}
\newcommand \eed {\end{displaymath}}

\newcommand{\bit}{\begin{itemize}}
\newcommand{\eit}{\end{itemize}}

\newcommand{\bgar}{\begin{eqnarray}}
\newcommand{\enar}{\end{eqnarray}}

\begin{document}

\title{Data analysis on Coronavirus spreading by macroscopic growth laws}

\author{P.~Castorina$^{(a,b)}$, A.~Iorio$^{(b)}$ and D.~Lanteri$^{(a,b,c)}$}
\affiliation{
\mbox{$^{(a)}$ INFN, Sezione di Catania, I-95123 Catania, Italy} \\
\mbox{$^{(b)}$ Faculty of Mathematics and Physics, Charles University} \\
\mbox{V Hole\v{s}ovi\v{c}k\'ach 2, 18000 Prague 8, Czech Republic} \\
\mbox{$^{(c)}$ Dipartimento di Fisica e Astronomia, Universit\`a di Catania, Italy}
}

\date{\today}
\begin{abstract}
\noindent To evaluate the effectiveness of the containment on the epidemic spreading of the new Coronavirus disease 2019, we carry on an analysis of the time evolution of the infection in a selected number of different Countries, by considering  well-known macroscopic growth laws, the Gompertz law, and the logistic law. We also propose here a generalization of Gompertz law. Our data analysis permits an evaluation of the maximum number of infected individuals. The daily data must be compared with the obtained fits, to verify if the spreading is under control. From our analysis it appears that the spreading reached saturation in China, due to the strong containment policy of the national government. In Singapore a large growth rate, recently observed, suggests the start of a new strong spreading. For South Korea and Italy, instead, the next data on new infections will be crucial to understand if the saturation will be reached for lower or higher numbers of infected individuals.
\end{abstract}
 \maketitle

\section{Introduction}

The epidemic spreading of the new Coronavirus disease 2019 (COVID-19)~\cite{oms} is producing the strongest containment attempt in recent history.
In many Countries world-wide millions of people are forced to live in isolation and in difficult conditions. In this work we focus on the Countries that first experienced the pandemic, that are China (where the infection appears now under control), and then South-Korea, Singapore, Italy (that show different degrees of containment effectiveness).

Since the mechanisms of COVID-19 spreading are not completely understood, the number of infected people is large, and the effects of containment are evaluated essentially on an  empirical basis. Therefore, a more quantitative analysis of the epidemic spreading  can be interesting. In the literature there is a large number of mathematical models (see for example~\cite{napoco1,napoco2,napoco3,napoco4}).

\begin{figure}
  \centering
  \includegraphics[width=0.45\textwidth]{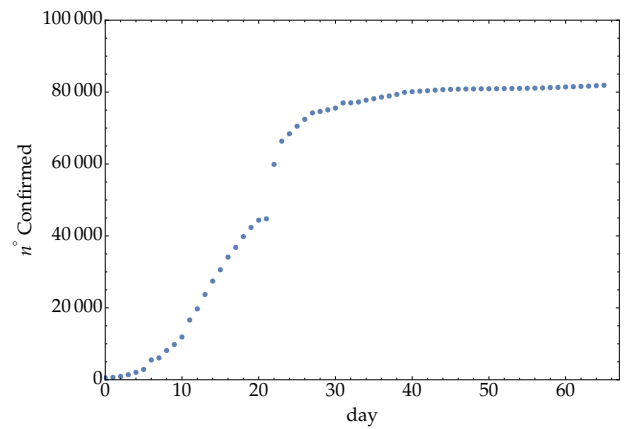}
  \caption{Number of infected individuals in China~\cite{hopkins}. The jump corresponds to a different counting rule of infected people.  Day zero is January the 22nd.}
  \label{figuno}
\end{figure}

\begin{figure}
  \centering
  \includegraphics[width=0.45\textwidth]{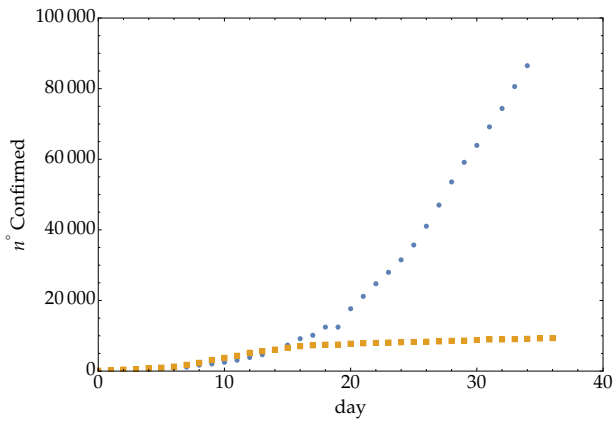}
  \caption{Number of infected individuals in South Korea (orange point) and Italy (blue point)~\cite{hopkins}.  Day zero is February
  	the 20th for South Korea, and February the 22nd for Italy.}
  \label{figdue}
\end{figure}

However, in our opinion, this stage of the disease does not permit a detailed comparative analyses, since the available data consist of the number of infected patients in different geographic areas, as shown in Fig.~\ref{figuno} for China and in Fig.~\ref{figdue} for South Korea and Italy~\cite{hopkins}, with different social, political and economical structures.

In other words, one has ``coarse-grained'' information and detailed ``microscopic'' studies that are, at the moment, of limited use since they have a larger number of free parameters with respect to macroscopic approaches, difficult to determine in a reliable way.

Moreover, there is an impressive number of experimental verifications, in many different scientific sectors, that  coarse-grain properties of systems, with simple laws with respect to the fundamental microscopic algorithms, emerge at different levels of magnification providing important tools for explaining and predicting new phenomena.

Therefore, an analysis based on macroscopic  laws can be useful to understand the behavior of growth rate of the infection and  to verify if its containment is indeed working.

A general classification of macroscopic growth laws is reported in Refs.~\cite{noi1,noi2}. In the present study we focus on  well-known laws: the Gompertz law (GL)~\cite{gompertz}, a new proposed  generalized GL (GGL) and the logistic law (LL)~\cite{logistic}, which will be compared with the exponential spreading, which means that the containment efforts have no effect.

The GL~\cite{gompertz}, initially applied to human mortality tables (i.e. aging) and  tumor growth~\cite{steel,norton}, also describes kinetics of enzymatic reactions, oxygenation of hemoglobin, intensity of photosynthesis as a function of CO2 concentration, drug dose-response curve, dynamics of growth, (e.g., bacteria, normal
eukaryotic organisms). The GGL is the generalization of the GL.

The LL~\cite{logistic} has been applied in population dynamics, in economics, in material science and in many other sectors.

The previous laws differ in the description of the virus containment effects, which in the LL is stronger (power law behavior) than in the GL and GGL, which have a logarithmic decrease of the specific growth rate (see appendix A).

For a discussion of the COVID-19 data, one has to know that each of the considered macroscopic laws is characterized by two important parameters, $\alpha_g$, $N_\infty^g$, for the GL, $\alpha_l$, $N_\infty^l$ for the LL, and by three parameters, $\alpha_{gg}$, $N_\infty^{gg}$ and $\beta$ for the GGL (the mathematical details are reported in appendix A). The meaning of the parameters is crucial to understand the evolution of the epidemic spreading.

The parameters $\alpha_g$, $\alpha_{gl}$, $\alpha_l$ describe the specific rate of the initial exponential growth, after which there is a slowdown of the disease, due to contrast mechanisms. In particular, $N_\infty^g$, $N_\infty^{gl}$, $N_\infty^l$, called \textit{carrying capacities}, fix the maximum number of infected people in the  models.

The \textit{contrast effect}, mathematically represented by the second term in Eqs.~(\ref{eq:1}),  (\ref{eq:2}) and (\ref{eq:3}), depends on many possible mechanisms of pathological and political origin (medical cure, biological conditions, isolation, information, \textit{et cetera}).

It should be clear that the present analysis does not give any specific indication in this respect, however the fitted value of  $N_\infty^g$, $N_\infty^{gl}$ and $N_\infty^l$ tell us how far is the disease evolution from the saturation point where the restraint effort is such that the spreading is practically over.

Indeed, a fit of the available data by GL, GGL and LL determines the values of the corresponding parameters, giving information on the possible behavior of the spreading, although the total number of infected people remains unknown, due to the large number of (a) untested, (b) paucisymptomatic and (b) fully asymptomatic population.

We apply the analysis to China, South Korea, Italy and Singapore since one needs the number on infected people in a large enough time interval for a reliable fitting procedure. Furthermore, those Countries are at sharply different stages of the spreading, with China essentially out of the emergency, South Corea half a way, and Italy still fully into it.

\section{Data analysis\label{sec:II}}

The cumulative number of infected people, in the different Countries, is used to describe the evolution of the infection spreading. However, the reliability of the data could  depend on the status of the spreading also: in China, where the counting of the infected people has been going on for a long time, the data are stable, and one does not expect
any systematic error due to external limiting factors.

On the other hand, in Italy,  not dynamical factors could reduce the effective number of infected people. If, for example, the number of available kits (swabs) to detected the disease has a maximum number per day, one cannot detected, in a single day, a larger number of infected individuals~\cite{parisi1}.

Moreover the asymptomatic population is unknown and any estimate is strongly model dependent \cite{lancet, istat, noi}. It should be clear that the variable $N(t)$ does not describe the total infected population but its time dependence includes, in an effective way, the dynamics among symptomatic and asymptomatic individuals \cite{istat,noi} and therefore it is a useful variable to understand the phase of the spreading.

With the previous warnings, in the next sections the global data about the cumulative number of infected people is discussed, for the different Countries, and compared with the macroscopic and exponential growth laws.

\section{How to use the fits}

To avoid possible misunderstandings, it is useful to comment on how to use of the previous fits in the future estimate.

With the caveats discussed in the previous section, the parameters $N_\infty$s give information on the maximum number of infected individuals. For each nation, one has to follow the day by day data, refitting the parameters until they stabilize. The key point is to check whether the data are in agreement either with the exponential, GL, GGL  and LL, or else are in between.
A typical example is given in table~\ref{tab:ita} where is reported for Italy the predictions obtained by using the available data until a specific day (March the 8th, for table~\ref{tab:ita}). The day after, one has to repeat the numerical analysis, which implies a redefinition of the parameters, i.e. of the specific growth rate, until they stabilize.

This is highly relevant, because the GL, GGL, describing a less effective containment effort, predict a much larger maximum number of infected. Hence, in this case, the contrast effort has to be improved and, probably, diversified. On the other hand, one gets a very good signal that the disease is slowing down to a smaller saturation values, if the data agree, or are less than, the values predicted by the LL, as in China.

\section{China}

\begin{figure}
  \centering
  \includegraphics[width=0.45\textwidth]{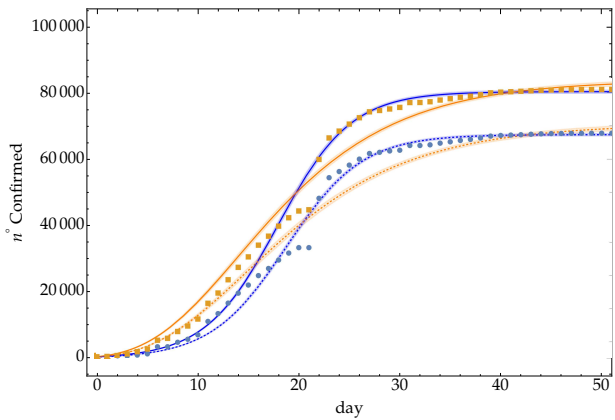}
  \caption{Number of infected individuals, fit of the Chinese data (orange points). The Gompertz law prediction (orange)  and the logistic curve (blue) are depicted with a band representing the $68\%$ of confidence level. The blue points and dotted curves describe to the same fit for the Hubei region. Time zero corresponds to the initial day - 22/01.}
  \label{figtre}
\end{figure}

The available data cover the long period from January the 22nd to March the 27th and, therefore, the numerical fit is more reliable.
The results are depicted in Fig.~\ref{figtre}. In the final period, when Chinese Government decided a different counting rule, the available data are well fitted by the
logistic curve with $\alpha_l = 0.278 \pm 0.003$ (per day) and $N_\infty^l= 80276 \pm 481$. Gompertz law predicts a larger saturation value $N_\infty^g = 83728 \pm 721$, with
$\alpha_g= 0.115 \pm 0.002$ (per day). Notice that the error is small due to the large number of available data.
The number of infected Chinese is today about $81897$, which means that the effort to contrast the disease has been successful and almost completed.
Previous analysis have been done by considering the growth of mortality~\cite{parisi1}.

\section{Singapore}
\begin{figure}
  \centering
  \includegraphics[width=0.45\textwidth]{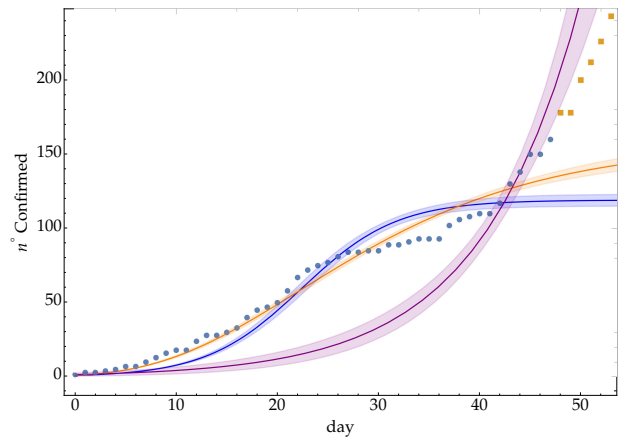}
  \caption{Number of infected individuals, fit of Singapore data. GL (orange) and LL (blue) with a band representing the 68 $\%$ of confidence level. Time zero corresponds to the initial day - 23/01. }
  \label{figsei}
\end{figure}

For Singapore, until March the 10th, the number of infected people is much smaller and the previous considered  external limiting factor does not, presumably, apply.
The resulting fit is  depicted in Fig.~\ref{figsei} and, as shown by data, there is a recent strong increase in the growth rate: a clear signal that there is some new uncontrolled outbreak of the infection with a related exponential trend.

\section{South Korea}

\begin{figure}
  \centering
  \includegraphics[width=0.45\textwidth]{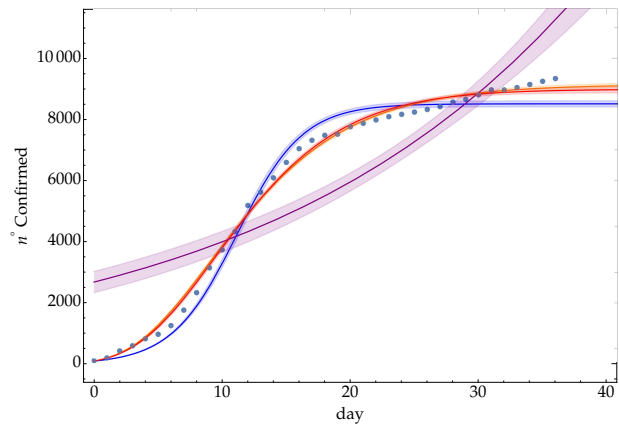}
  \caption{Number of infected individuals, fit of the South Korea data. GL (orange), GGL (red), LL (blue) and Exponential (purple), with a band representing the $68 \%$ of confidence level. Time zero corresponds to the initial day - 20/02.}
  \label{figquattro}
\end{figure}

Fig.~\ref{figquattro} shows the result of the fits using the South Korea data, from February the 20th to March the 27th. The reduced number of data increases the error in the fitted parameter:
$\alpha_g= 0.165 \pm 0.003$, $\alpha_l= 0.393 \pm 0.007$, $N_\infty^g = 9145 \pm 78$ and $N_\infty^l =8506 \pm 98$.
The GGL parameters are $\alpha_{gg}=0.172\pm0.005$, $N^{gg}_\infty=8976\pm112$ and $\beta = 0.06\pm0.04$

The Gl and LL differ in the saturation values, although they are compatible within the $68\%$ of confidence level (see the band in Fig.~\ref{figtre}). Therefore one has to carefully follow if the next data are in agreement with the Gompertz evolution or with the logistic one. The exponential behavior is strongly disfavored by the data.

\begin{figure}
  \centering
  \includegraphics[width=0.45\textwidth]{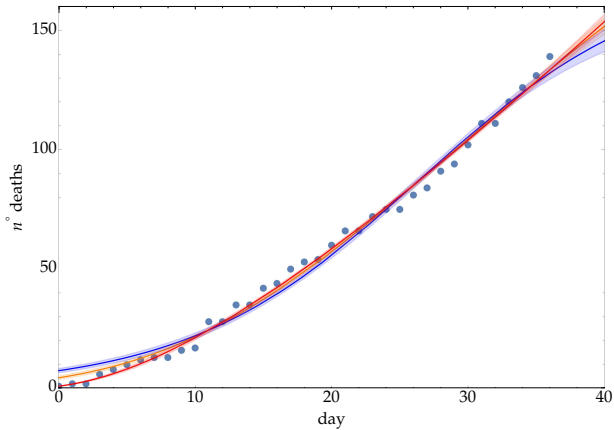}
  \caption{Mortality growth, fit of the South Korea data. GL (orange), GGL (red), LL (blue). Time zero corresponds to the initial day - 20/02.}
  \label{figquattrobis}
\end{figure}

The mortality growth follows the same trend, as shown in Fig.~\ref{figquattrobis}.

\section{Italy}

\begin{figure}
  \centering
  \includegraphics[width=0.45\textwidth]{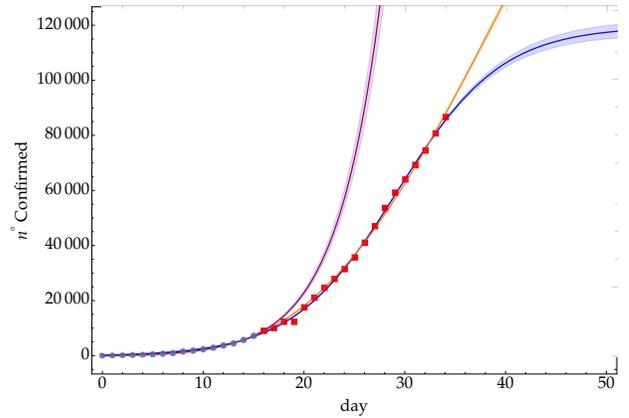}
  \caption{Number of infected individuals, fit of Italy data. Exponential law (purple), GL (orange) and LL (blue) with a band representing the 68 $\%$ of confidence level. Time zero corresponds to the initial day - 22/02. The parameters are:  $\alpha_g= 0.053 \pm 0.001$, $\alpha_l= 0.194 \pm 0.003$, $N_\infty^g = 364431 \pm 16385$ and $N_\infty^l = 119238 \pm 2630$.
The GGL gives results very similar to the GL. Red data have not been included in the exponential fit.}
  \label{figcinque}
\end{figure}

The Italian data cover the time range going from February the 22nd to March the 27th. The results are depicted in Fig.~\ref{figcinque}, where the band represent the $68\%$ of confidence level. Previous analysis has been done in ref.~\cite{marinari}, looking at the mortality table and at the number of patients in the Italian hospitals.
The data in Fig.~\ref{figcinque} are well fitted by the GL and the LL is plotted to verify the signal of a stronger reduction in a possible saturation phase, as observed in Chinese data.

The initially large specific rate forces the GL to reproduce the data with an artificially large value of $N_\infty^{gl}$, due to the logarithmic behavior, but with a large error. The deviation of GL and LL from the exponential growth is more readable in Table~\ref{tab:ita} and the exponential growth predicts much larger value of cumulative infected individuals in the next days.

\begin{figure}
  \centering
  \includegraphics[width=0.45\textwidth]{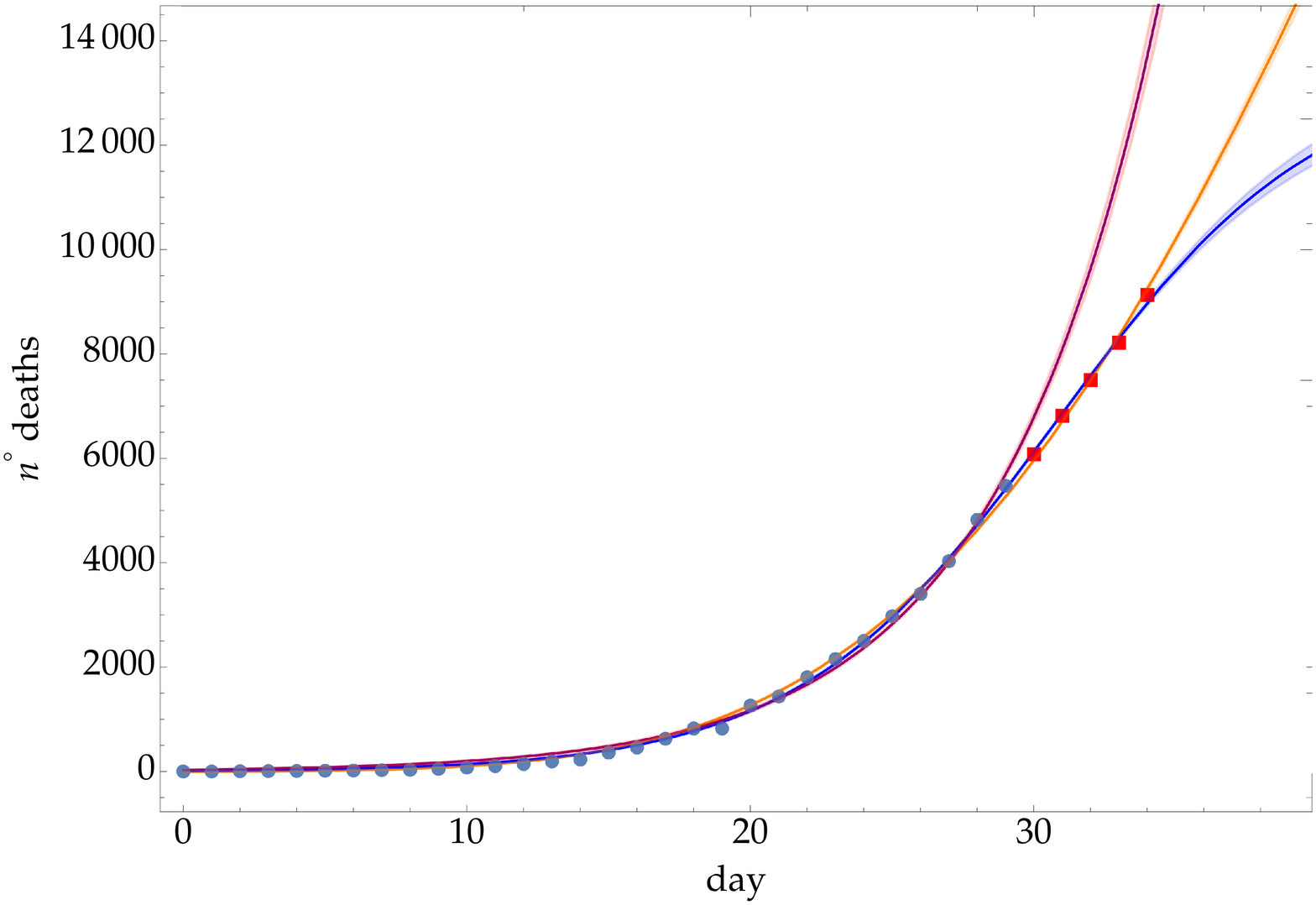}
  \caption{Mortality growth, fit of the Italian data by exponential law (purple), GL (orange) and LL (blue). Time zero corresponds to the initial day - 20/02.}
  \label{figquattroter}
\end{figure}

The mortality follows a similar trend until about March the 22nd, see Fig.~\ref{figquattroter}, with some delay with respect to the $N(t)$ behavior.

Finally, the values of the fitted parameters are summarized in tables~\ref{tab:1} and~\ref{tab:2} for a comparison between different nations.

\begin{table}
	\centering 	
	\begin{tabular}{|c|c|c|c|c|}
		\hline
		& $\alpha_g$ & $\alpha_l$ \\
		\hline
		\text{China} & $0.115\pm 0.002$ & $0.278\pm 0.003$ \\
		\text{South Korea} & $0.165\pm 0.003$ & $0.393\pm 0.007$ \\
		\text{Italy} & $0.053\pm 0.001$ & $0.194\pm 0.003$ \\
		\text{Singapore} & $0.073\pm 0.003$ & $0.213\pm 0.007$ \\
		\hline
	\end{tabular}
	\caption{The value of the parameters $\alpha_g$ and $\alpha_l$ for different Nations. }
	\label{tab:1}
\end{table}

\begin{table}
	\centering
	\begin{tabular}{|c|c|c|}
		\hline
		& $N_{\infty }^g$ & $N_{\infty }^l$ \\
		\hline
		\text{China} & $83728\pm 721$ & $80276\pm 481$ \\
		\text{South Korea} & $9145\pm 78$ & $8506\pm 98$ \\
		\text{Italy} & $364431\pm 16385$ & $119238\pm 2630$ \\
		\text{Singapore} & $158\pm 7$ & $119\pm 4$ \\
		\hline
	\end{tabular}
	\caption{The maximum number of infected individuals evaluated by the fitting procedure in different Countries.}
	\label{tab:2}
\end{table}

\begin{table}
	\centering
	\begin{tabular}{|c|c|c|c|c|}
		\hline 	
		day (March) & N & Exp & GL & LL \\
		\hline 	
		1  \text{th} & 1694 & 1512 & 1255 & 1912 \\
		2 \text{th} & 2036 & 1898 & 1684 & 2314 \\
		3 \text{th} & 2502 & 2383 & 2225 & 2798 \\
		4 \text{th} & 3089 & 2991 & 2898 & 3381 \\
		5 \text{th} & 3858 & 3755 & 3723 & 4081 \\
		6 \text{th} & 4636 & 4713 & 4720 & 4919 \\
		7 \text{th} & 5883 & 5916 & 5913 & 5920 \\
		8 \text{th} & 7375 & 7425 & 7320 & 7113 \\
		\hline
		9 \text{th} & 9172 & 9320 & 8962 & 8528 \\
		10 \text{th} & 10149 & 11699 & 10859 & 10199 \\
		11 \text{th} & 12462 & 14685 & 13027 & 12161 \\
		12 \text{th} & 12462 & 18433 & 15481 & 14451 \\
		13 \text{th} & 17660 & 23137 & 18233 & 17103 \\
		14 \text{th} & 21157 & 29042 & 21293 & 20148 \\
		15 \text{th} & 24747 & 36455 & 24668 & 23610 \\
		16 \text{th} & 27980 & 45759 & 28361 & 27501 \\
		17 \text{th} & 31506 & 57437 & 32372 & 31821 \\
		18 \text{th} & 35713 & 72096 & 36698 & 36549 \\
		19 \text{th} & 41035 & 90496 & 41332 & 41645 \\
		20 \text{th} & 47021 & 113593 & 46266 & 47047 \\
		21 \text{th} & 53578 & 142584 & 51487 & 52674 \\
		22 \text{th} & 59138 & 178974 & 56981 & 58429 \\
		23 \text{th} & 63927 & 224652 & 62731 & 64206 \\
		24 \text{th} & 69176 & 281988 & 68718 & 69899 \\
		25 \text{th} & 74386 & 353957 & 74921 & 75404 \\
		26 \text{th} & 80589 & 444294 & 81319 & 80634 \\
		27 \text{th} & 86498 & 557686 & 87890 & 85520 \\
		28 \text{th} & \text{} & 700019 & 94611 & 90011 \\
		29 \text{th} & \text{} & 878678 & 101457 & 94080 \\
		30 \text{th} & \text{} & 1102934 & 108406 & 97718 \\
		31 \text{th} & \text{} & 1384424 & 115433 & 100932 \\
		\hline 	
	\end{tabular}
	\caption{Number of confirmed sick in Italy predicted by exponential, Gompertz and logistic fits, compared with data (column N). Fits are made by using the available data until March the 8th for exponential grow, until March the 25 for the gompertz and logistic ones.}
	\label{tab:ita}
\end{table}

\section{Comments and conclusions}

Let us state clearly that, the take-home message of our analysis is that, beyond any doubts, a strong containment policy should be kept.

As for countries with a longer (known) exposure to COVID-19, our analysis clearly shows that the spreading: a) has reached saturation in China, b) but in Singapore, after a period of important slow down, a new increase is clearly visible. As for countries with a shorter (known) exposure, keeping in mind the limitations recalled in Sec.~\ref{sec:II}, our analysis, depicted in Figs.~\ref{figquattro} and~\ref{figcinque}, shows that South Korea and Italy are in different situations (see also \cite{chen,giudici}). In Italy, the observed data in the near future will be crucial to understand if the evolution will either follow an exponential growth, or the GL, or the GGL or else the LL. This will allow to understand if the saturation will be reached for lower or higher numbers of infected individuals. The proposed approach for monitoring the evolution of the epidemic spreading of COVID-19 has to be consider as a complementary tool to more fundamental genomics methods~\cite{ita1}.

Of course, this analysis needs to be updated on a daily basis. The daily data must be compared with the fits, to verify if the spreading is under control or not (out of control being the exponential growth). This will help to understand quantitatively the status of the COVID-19 spreading.

\section*{Acknowledgments}

The authors thank Giorgio Parisi for useful discussions and comments. A.I. is partially supported by UNCE/SCI/013.

\appendix

\section{\label{app:A}}

Let us call $N(t)$ the number of infected individuals at time $t$. The Gompertz evolution law is the solution of the differential equation
\begin{equation}\label{eq:1}
\frac{1}{N(t)}\;\frac{dN(t)}{dt} = \alpha_g\;\ln \frac{N_\infty^g}{N(t)} \,,
\end{equation}
the Generalized Gompertz law is solution of
\begin{equation}\label{eq:2}
\frac{1}{N(t)}\;\frac{dN(t)}{dt} = \alpha_{gg}\; ln ^{\left(1-\beta\right)}\left(\frac{N_\infty^g}{N(t)}\right) \,,
\end{equation}
while the logistic law equation is
\begin{equation}\label{eq:3}
\frac{1}{N(t)}\;\frac{dN(t)}{dt} = \alpha_l\;\left( 1 - \frac{N(t)}{N_\infty^l}\right) \,.
\end{equation}
The exponential behaviour (i.e. no reduction of the spreading) is
\begin{equation}\label{eq:4}
\frac{1}{N(t)}\;\frac{dN(t)}{dt} = constant \,,
\end{equation}
The  laws differ in the description of the contrast term in the second member.

The general solution of the Gompertz equation is
\begin{equation}\label{eq:5}
N^g(t) = N_\infty^g\; \exp\left\{\ln\left(\frac{N(0)}{N_\infty^g}\right)\;e^{-\alpha_g (t-t_0)}\right\}
\end{equation}
where $t_0$ is the initial time, $N(0)$ the initial value of the infected individuals coming from the available data. The generalized Gompertz solution is
\begin{equation}\label{eq:6}
N^{gg}(t) = N_\infty^{gg}\;\exp\left\{-\left[\ln^{\textstyle{\beta}}\left(\frac{N_\infty^{gg}}{N(0)}\right)-\alpha_{gg}\;\beta\;(t-t_0)\right]^{\frac{1}{\beta}}\right\}
\;.
\end{equation}
For the logistic equation one has
\begin{equation}\label{eq:7}
N^l(t)= \frac{N(0)\;e^{\alpha_l(t-t_0)}}{1 - \frac{N(0)}{N_\infty^l}\;\left[1 - e^{\alpha_l(t-t_0)}\right]} \,.
\end{equation}


\begin{thebibliography}{99}
\bibitem{oms} World Health Organization, Coronavirus disease (COVID-19) outbreak, \\
https://www.who.int/emergencies/diseases/novel-coronavirus-2019.
\bibitem{napoco1} S.~A. Herzog, S.~Blaizot and Niel Hens, Mathematical models used to inform study design or surveillance systems in infectious diseases: a systematic review, BMC Infectious Diseases  \textbf{17} (2017) 775.
\bibitem{napoco2}  N.~C.~Grassly and C.~Fraser, Mathematical models of infectious disease transmission, Nature Reviews Microbiology \textbf{6} (2008) 477.
\bibitem{napoco3}  R. Pastor-Satorras, C. Castellano, P.Van Mieghem and A.Vespignani , Epidemic processes in complex network, Rev. Mod. Phys., VOLUME 87 (2015).
\bibitem{napoco4} P.Blanchard, G.F. Bolz and T.Kruger, Mathematical modelling on random graphs of sesually trasmitted disease, in Dynamics and Stochastic Process - Theory and Applications, Lecture Notes in Physics, vol. 355, Springer-Verlag, Berlin.
\bibitem{hopkins}  Novel Coronavirus (COVID-19) Cases, provided by JHU CSSE,  https://github.com/CSSEGISandData/COVID-19.
\bibitem{noi1}  P.~Castorina, P.~P.~Delsanto, C.~Guiot, Classification Scheme for Phenomenological Universalities
   in Growth Problems in Physics and Other Sciences, Phys. Rev. Lett. \textbf{96} (2006) 188701.
\bibitem{noi2} P.~Castorina and P.~Blanchard, Unified approach to growth and aging in biological, technical and biotechnical systems, SpringerPlus \textbf{1} (2012) 7.
\bibitem{gompertz}  B. Gompertz, On the nature of the function expressive of the law of human mortality and a new mode of determining
life contingencies, Phil. Trans. R. Soc. \textbf{115} (1825) 513.
\bibitem{logistic}  P.~F.~Verhulst, Notice sur la loi que la population poursuit dans son accroissement, Correspondance Math\'ematique et Physique, \textbf{10} (1838) 113.
\bibitem{parisi1} G. Parisi, private communication.
\bibitem{lancet} A.R.Tuite, V. Ng, E. Rees and D.Fisman,  Estimation of COVID-19 outbreak size in Italy, The Lancet Infec. Dis. March 19, doi.org/10.1016/S1473-3099(20)30227-9.
\bibitem{istat} L.Fenga,  CoViD19: An Automatic,Semiparametric Estimation Method for the Population Infected in Italy, medRxiv preprint doi: https://doi.org/10.1101/2020.03.14.20036103.
\bibitem{noi} D.~Lanteri, D.~Carco' and P.~Castorina, How macroscopic laws describe complex dynamics: asymptomatic population and CoviD-19 spreading, arXiv:2003.12457.
\bibitem{who} World Health Organization, Coronavirus disease (COVID-19) report, https://www.who.int/docs/default-source/coronaviruse/who-china-joint-mission-on-covid-19-final-report.pdf
\bibitem{marinari} E. Bucci, E. Marinari, L$'$evoluzione dell$'$epidemia da coronavirus in Italia, Scienza in Rete (in Italian) https://www.scienzainrete.it/
\bibitem{steel} G.G: Steel, Growth kinetics of tumours, Clarendon Press, Oxford, 1977.
\bibitem{norton} L.~A.~Norton, Gompertzian model of human breast cancer growth, Cancer. Res. \textbf{48} (1988) 7067.
\bibitem{chen} Y. Chen, Q. Liu, D. Guo, Emerging coronaviruses: Genome structure, replication, and pathogenesis, J. Med. Virol. \textbf{92} (2020) 418.
\bibitem{giudici}  A.Arianna and P. Giudici,  A Poisson Autoregressive Model to Understand COVID-19 Contagion Dynamics (March 9, 2020). Available at SSRN: https://ssrn.com/abstract=3551626 or http://dx.doi.org/10.2139/ssrn.3551626
\bibitem{ita1} A. Lai,  A. Bergna, C. Acciarri, M. Galli, G. Zehender, Early Phylogenetic Estimate of the Effective Reproduction Number Of Sars-CoV-2, J. Med. Virol. 2020 Feb 25, doi: 10.1002/jmv.25723.
\end{thebibliography}
\end{document}